\documentclass[aps,prl,twocolumn,superscriptaddress,showpacs,longbibliography]{revtex4-1}
\usepackage[colorlinks=true,citecolor=blue,linkcolor=blue,breaklinks=true]{hyperref}

\usepackage{color,graphicx}
\usepackage{bm}
\usepackage{amsmath}
\usepackage{amssymb}

\def\nd{^{\vphantom{\dagger}}}
\def\yd{^\dagger}
\def\frac#1#2{{\textstyle{#1 \over #2}}}

\def\CM{{\cal M}}
\def\CH{{\cal H}}
\def\CO{{\cal O}}

\def\CK{{\cal K}}
\def\CR{{\cal R}^{\rm corr}}
\def\ve{\varepsilon}
\def\bk{{\boldsymbol k}}

\def\bh{{\boldsymbol h}}

\def\bsigma{{\boldsymbol\sigma}}
\def\vek{\ve(\bk)}
\def\veki{\ve\nd_i(\bk)}
\def\veF{\ve_{\rm F}}
\def\kF{k_{\rm F}}
\def\sgn{\textsf{sgn}}
\def\etal{{\it et al.\/}}

\def\sssf#1{{\scriptscriptstyle\textsf{#1}}}
\def\inter{\sssf{INTER}}
\def\intra{\sssf{INTRA}}
\def\cmc{\sssf{CMC}}
\def\csr{\sssf{CSR}}
\def\ccmc{\chi\nd_\cmc}
\def\ccsr{\chi\nd_\csr}
\def\pz{\partial}
\def\xhat{{\hat{\boldsymbol x}}}
\def\yhat{{\hat{\boldsymbol y}}}
\def\zhat{{\hat{\boldsymbol z}}}
\def\tdc{\textsf{3dC}}
\def\tdh{\textsf{2dH}}
\def\tj{{j^\alpha_\intra}}
\def\tM{{M_\intra}}

\def\be{\begin{equation}}
\def\ee{\end{equation}}

\def \be{\begin{equation}}
	\def \ee{\end{equation}}
\def \bea{\begin{eqnarray}}
	\def \eea{\end{eqnarray}}

\def\zhat{\hat{\bf z}}

\def\cH{{\cal H}}

\def\frac#1#2{{\textstyle{#1 \over #2}}}

\def\cO{{\cal O}}

\def\ve{\varepsilon}

\def\Im{{\rm Im}}

\mathchardef\Gamma="7100

\newcommand {\no} {\nonumber}

\newcommand{\lt} {\left}
\newcommand{\rt} {\right}

\def\bk{{\bf k}}

\def\bh{{\bf h}}
\def\vsigma{\mbox{\boldmath{$\sigma$}}}
\def\tbphi{\tilde{\mbox{\boldmath{$\phi$}}}}

\def\ek{{\epsilon_\bk}}

\def\tphi{\tilde{\phi}}
\def\ttheta{\tilde{\theta}}

\def\tj{{\tilde j}}

\def\hj{{\hat{ j}}}
\def\tM{{\tilde M}}

\def\ek{{\epsilon_\bk}}
\def\Tr{{\rm Tr}}

\def\nd{^{\vphantom{\dagger}}}
\def\yd{^\dagger}
\def\frac#1#2{{\textstyle{#1 \over #2}}}

\def\CM{{\cal M}}
\def\CH{{\cal H}}
\def\CO{{\cal O}}

\def\CK{{\cal K}}
\def\ve{\varepsilon}
\def\bk{{\boldsymbol k}}

\def\bh{{\boldsymbol h}}
\def\vsigma{{\boldsymbol\sigma}}
\def\vek{\ve(\bk)}
\def\veki{\ve\nd_i(\bk)}
\def\veF{\ve_{\rm F}}
\def\kF{k_{\rm F}}
\def\sgn{\textsf{sgn}}
\def\etal{{\it et al.\/}}

\def\sssf#1{{\scriptscriptstyle\textsf{#1}}}
\def\ccmc{\chi\nd_\ssf{CMC}}
\def\ccsr{\chi\nd_\ssf{CSR}}
\def\pz{\partial}
\def\xhat{{\hat{\boldsymbol x}}}
\def\yhat{{\hat{\boldsymbol y}}}
\def\zhat{{\hat{\boldsymbol z}}}
\def\inter{\sssf{INTER}}
\def\intra{\sssf{INTRA}}
\def\tdc{\textsf{3dC}}
\def\tdh{\textsf{2dH}}
\def\tj{{j^\alpha_\intra}}
\def\tM{{M_\intra}}
\def\hj{{\hat j}}
\def\hA{{\hat A}}
\def\hP{{\hat P}}
\def\nd{^{\vphantom{\dagger}}}
\def\yd{^\dagger}
\def\frac#1#2{{\textstyle{#1 \over #2}}}

\def\CM{{\cal M}}
\def\CH{{\cal H}}
\def\CO{{\cal O}}

\def\CK{{\cal K}}
\def\ve{\varepsilon}
\def\bk{{\boldsymbol k}}

\def\bh{{\boldsymbol h}}
\def\vsigma{{\boldsymbol\sigma}}
\def\vek{\ve(\bk)}
\def\veki{\ve\nd_i(\bk)}
\def\veF{\ve_{\rm F}}
\def\kF{k_{\rm F}}
\def\sgn{\textsf{sgn}}
\def\etal{{\it et al.\/}}

\def\sssf#1{{\scriptscriptstyle\textsf{#1}}}
\def\inter{\sssf{INTER}}
\def\intra{\sssf{INTRA}}
\def\cmc{\sssf{CMC}}
\def\csr{\sssf{CSR}}
\def\ccmc{\chi\nd_\cmc}
\def\ccsr{\chi\nd_\csr}
\def\pz{\partial}
\def\xhat{{\hat{\boldsymbol x}}}
\def\yhat{{\hat{\boldsymbol y}}}
\def\zhat{{\hat{\boldsymbol z}}}
\def\tdc{\textsf{3dC}}
\def\tdh{\textsf{2dH}}
\def\tj{{j^\alpha_\intra}}
\def\tM{{M_\intra}}

\begin{document}
\title{Hall coefficient of  semimetals}
\author{Abhisek Samanta}\email{abhiseks@campus.technion.ac.il}
\affiliation{ Physics Department, Technion, Haifa 32000, Israel}
\author{Daniel P. Arovas}\email{darovas@ucsd.edu}
\affiliation{ Department of Physics, University of California at San Diego, La Jolla, California 92093, USA}
\author{Assa Auerbach}\email{assa@physics.technion.ac.il}
\affiliation{ Physics Department, Technion, Haifa 32000, Israel}

\date{\today }
\begin{abstract}
	A recently developed formula for  the Hall coefficient  [A. Auerbach,
	Phys. Rev. Lett. {\bf 121},  66601 (2018)]  is applied to nodal line and Weyl semimetals (including graphene), and to spin-orbit split semiconductor bands in two and three dimensions.
	The calculation reduces to a ratio of two equilibrium susceptibilities, where corrections are negligible at weak disorder. Deviations from Drude's inverse carrier density are associated with
	band degeneracies, Fermi surface topology, and interband currents. Experiments which can  measure these deviations are proposed.

\end{abstract}

\maketitle

Semimetals are characterized by proximity of the Fermi energy to band degeneracies. Vigorous research has been recently invested in semimetals on surfaces of
topological insulators~\cite{TI1,TI2}, Dirac and Weyl semimetals~\cite{Weyl1, Weyl2,Weyl3, NodalSM1, NodalSM2, NodalSM3}, and on semimetal platforms for Majorana states applications~\cite{Majorana}.

This paper focuses on the Hall coefficient of semimetals, which has been traditionally used to measure the charge carrier density $n$ using Drude's relation  $R_{\rm H} \propto n^{-1}$. In semimetals, Drude's relation may break down due to multiband effects, and  Fermi surface topology. For example, corrections to Drude's relation was found by Liu \etal  \cite{Culcer} for  spin-orbit split semiconductor bands. Multiband conductivity calculations involve coupled Boltzmann equations 
with interband collision integrals which are quite challenging \cite{Bulbul, Assa-Allen}.

We can avoid  coupled Boltzmann equations by applying the Hall coefficient formula \cite{PRL, EMT} to multiband Hamiltonians.
The dissipative scattering rates drop out,  and  
$R_{\rm H}$ is primarily determined by the non-dissipative Lorentz force captured by
the {\em current-magnetization-current} (CMC) susceptibility $\ccmc$,  and the {\em conductivity sum rule}  (CSR) $\ccsr$. {  Both coefficients are non dissipative: the CMC describes the effect of the Lorentz force on the currents, and the CSR governs their reactive response.}

Crucial to our approach  is the estimation of the formula's correction term ${\cal R}^{\rm corr}$, which  is determined by higher moments of the dynamical conductivity. 
This paper shows that  in the ``good quasiparticles'' (Boltzmann) regime,  ${\cal R}^{\rm corr}$ can be neglected for disorder strength less than the Fermi energy.   

\begin{figure}[t]
	\includegraphics[width=0.47\textwidth]{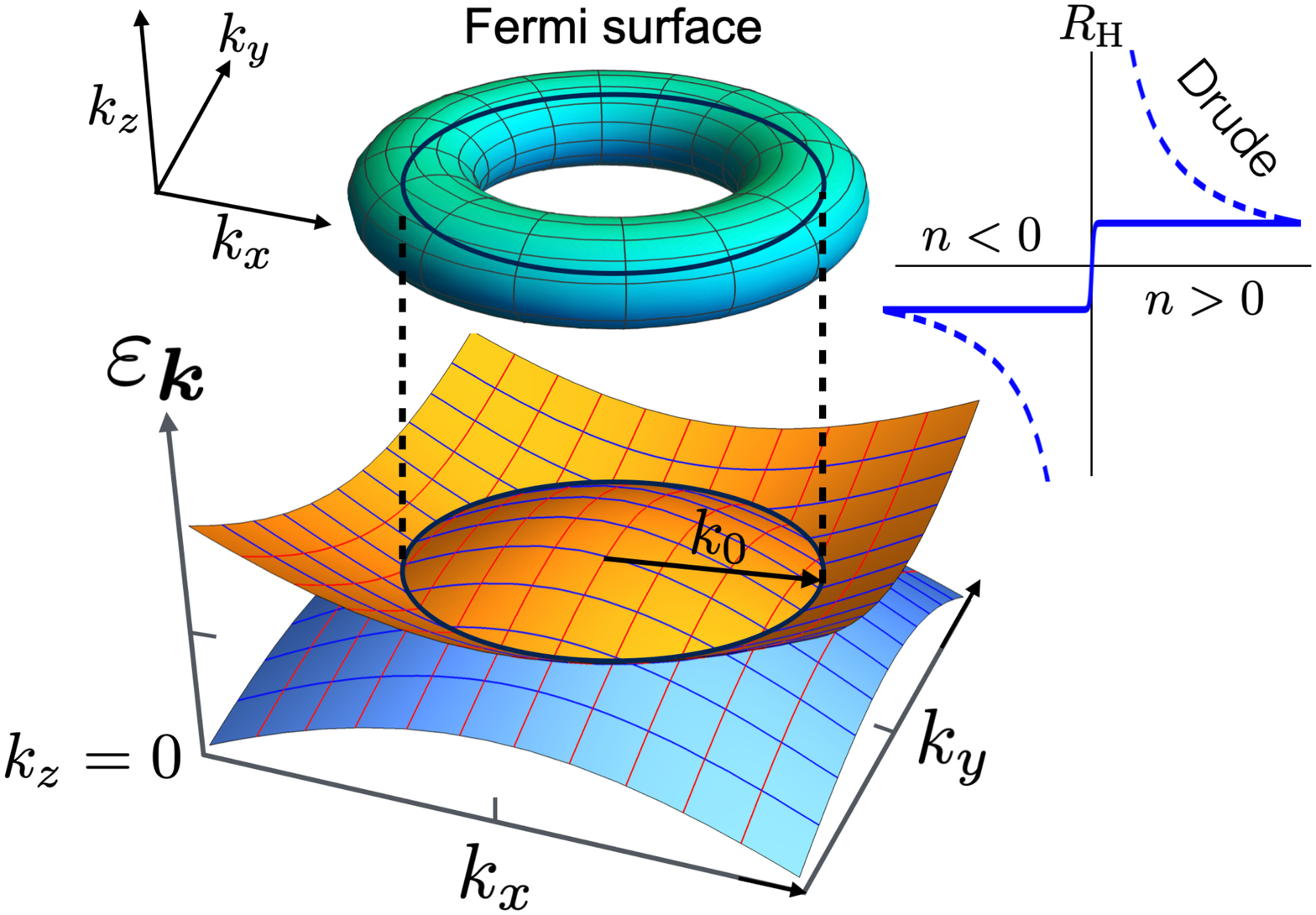}
	\caption{Nodal line semimetal.  The nodal line is marked by black circle  of radius $k_0$. The  three dimensional  toroidal Fermi surface (top) is depicted.
		At the upper right corner, the qualitative behavior of the Hall coefficient (solid line) is compared to Drude relation, for density $n$ as measured from the nodal circle filling.}
	\label{fig:NLSM}
\end{figure}

Our key results are:   (i) For Weyl  point semimetals in two and three dimensions, (including graphene) the intraband $R_{\rm H}^\intra(n)$ exhibits a Drude-like divergence, which is cut  off by interband scattering at low densities.
(ii) The nodal line semimetal (see Fig.~\ref{fig:NLSM}) exhibits a constant (rather than diverging) Hall coefficient, with a sign change
at the nodal energy. 
(iii) Previous results \cite{Culcer} of  spin-orbit split bands are extended into the interband transport regime, and to  three dimensions.
(iv) ${{\cal R}^{\rm corr}}$ is shown to be relatively suppressed by the disorder potential fluctuations divided by the Fermi energy squared.
The paper ends with a summary and proposals for experiments.

{\em The Hall coefficient formula },  as derived directly from the Kubo formulas~ \cite{PRL,EMT} for any Hamiltonian $\CH$ and spectrum $\{E_n,|n\rangle\}$,
\begin{equation}
R_{\rm H}  \equiv   \lim_{B\to 0}\bigg(\sigma_{xx}^{-2}\,{\pz\sigma_{xy}\over\pz B}\bigg)=  { \ccmc  \over   \chi^2_\csr } +\CR   .\label{RH-gen}
\end{equation}
$\sigma_{\alpha\beta}$  is the conductivity tensor (assuming C4 symmetry) and $B$ is the magnetic field in the $z$-direction. 
$\ccmc= \Im \big( j^y, [ M, j^x]\big) -\Im \big(j^x, [M, j^y]\big)$ and $\ccsr =  \big( j^x,j^x\big)$, where $j^\alpha$ is the uniform current in the $\alpha$ direction, and  $M= \pz\CH/\pz B$ is the diamagnetization operator.
{  The inner products are defined by equilibrium susceptibilities,
	\cite{SM},
	\begin{equation}
	(j^\alpha,A) = {1\over Z} \sum_{nm} {e^{-\beta E_n} - e^{\-\beta E_m} \over E_m-E_n } \langle n|  j^\alpha |m\rangle  \langle m| A |n\rangle\quad ,
	\end{equation}
	where $Z= {\rm Tr} e^{-\beta\CH}$ and  $\beta$ is the inverse temperature. }

The  correction term $\CR$ is given by
\begin{align}
\CR &= {1\over \ccsr}\sum_{i,j=0}^\infty (1-\delta\nd_{i,0}\,\delta\nd_{j,0})  \CM''_{2i,2j} \label{Rcorr}\\
&\qquad\times\prod_{i'=0}^i \left( -{\Delta_{2i'-1}\over \Delta_{2i'~~}}\right) \prod_{j'=0}^j \left( -{\Delta_{2j'-1}\over \Delta_{2j'~~}}\right) \quad.
\nonumber
\end{align}
$\CM''_{2i,2j}$ are cross-susceptibilities,  defined by the matrix elements of the magnetization commutator $[M,\bullet]$  between two currents'  Krylov bases.
The Krylov bases are generated by orthonormalizing the sets  of operators $ [\CH,[\>\cdots, [ \CH ,j^\alpha ] ] ]  $.  
$  \Delta_{i'} $  are the conductivity recurrents \cite{LA}, which can be obtained from the conductivity moments, defined by $\mu_{2i} =\langle \big[j^x, [\CH,[\>\cdots, [ \CH ,j^x ] ] ]\big]\rangle$,
where $\CH$ appears $2i-1$ times.  Instructions for calculating $\CM''_{2i,2j}$ and $\Delta_{i'}$  are reviewed in \cite{SM}.
{  Physically, $\CR$ incorporates the higher order effects of current non-conservation. In several examples, its relative magnitude can be suppressed  by using a renormalized Hamiltonian \cite{EMT}.}
Later, we  estimate $\CR$ and show that it can be neglected  in regimes of weak disorder which concern  this paper. 

We consider a general two-band Hamiltonian,
\begin{equation}
\CH_0 \equiv \sum_\bk\sum_{l,l'=1}^2 c\yd_{l\bk } \, h\nd_{ll' } (\bk) \, c\nd_{l'\bk}\quad.
\label{Hk}
\end{equation}
where $c\yd_{l\bk }$ creates a Bloch electron on band $l$ and wavevector $\bk$. A random potential with fluctuation $ V^2_{\rm dis}$  introduces a transport scattering rate $\hbar /\tau_{\rm tr} \sim V_{\rm dis}^2 /|\veF|$, 
{  where $\veF$ is the Fermi energy measured from the nearest particle-hole symmetric energy or band extremum.  }

Within the ``good quasiparticles'' regime, $\hbar/\tau_{\rm tr}\ll \veF\nd $,  the ratio
of the disorder strength to interband gap at the Fermi energy  $\Delta\ve$, defines two distinct transport regimes.
Importantly,  for evaluation of Eq.~(\ref{RH-gen}),  we have the freedom to choose the (renormalized) effective Hamiltonian which best describes the low energy correlations. Our choice determines the values of  $\ccmc,\ccsr$ and ${{\cal R}^{\rm corr}} $. It is the latter we wish to minimize.

{\em  (i)  Intraband regime} applies  for $V_{\rm dis}^2 \ll (\Delta\ve)^2 $, where
interband scattering is suppressed, and transport is dominated by  band-diagonal current and magnetization operators:
\begin{eqnarray}
\tj &=&  e \sum_{i,\bk}c\yd_{i\bk} \, v^\alpha_i(\bk) \, c\nd_{i\bk}~,~\alpha=x,y~,\quad i=1,2, \nonumber\\
\tM  &=& {ie\over 2c}  \sum_{i,\bk} c\yd_{i\bk}\, \bigg(\!v^y_i(\bk) {\pz\over\pz k_x} - v^x_i(\bk) {\pz\over\pz k_y}  \bigg) \, c\nd_{i\bk}
\label{intra-curr}
\end{eqnarray}
with $v^\alpha_i(\bk)  =\pz\veki/\pz k_\alpha$, where $\veki$ $(i\!=\!1,2)$ are the eigenvalues of $h_{ll'}(\bk)$. The
susceptibilities in this regime are \cite{SM},
\begin{equation}
\begin{split}
\chi_\cmc^\intra&={e^3\over c} \sum_{i=1}^2 \int \!{d^d k\over (2\pi)^d} \, F\nd_i(\bk) \,
\bigg(\!\!-{\pz f\over\pz\ve}\bigg)\nd_{\!\ve=\ve\nd_i(\bk)}\\
F\nd_i(\bk)&=  \big[v^y_i(\bk) \big]^{\!2} \,{\pz v^x_i(\bk)\over\pz k_x} - v^x_i(\bk) \, v^y_i(\bk) \, {\pz v^y_i(\bk)\over\pz k_x}\\
\chi_\csr^\intra  &= e^2  \sum_{i=1}^2\int\! {d^d k\over (2\pi)^d }\,\big(v^x_i(\bk) \big)^{\!2} 
\bigg(\!\!-{\pz f\over\pz\ve}\bigg)\nd_{\!\ve=\ve\nd_i(\bk)}\quad.
\label{csr-intra}
\end{split}
\end{equation}
$f_i $ is the Fermi-Dirac distribution of band $\veki$ at temperature $T$ and chemical potential $\veF$.  
For any spherically symmetric  band, $\vek=\ve(k)$, Drude's relation  holds, i.e. $R_{\rm H}=\ccmc/\chi_\csr^2=  1/(nec)$  \cite{Comm-elliptical}.
For more general band structures, Eqs.~(\ref{csr-intra}) recovers the venerable Boltzmann equation result in the  ``isotropic scattering limit'' \cite{Ziman, Comm-ISL}. 

{\em  (ii)  Interband regime} applies within the range  $(\Delta\ve)^2 \le  V_{\rm dis}^2 \ll \veF^2 $, where disorder is strong enough to mix the two bands (but still weak enough to neglect ${{\cal R}^{\rm corr}}$, see later discussion).
Interband currents now contribute to the longitudinal conductivity and to $ \ccsr$~\cite{Bulbul,Mohit}.  In this regime,  the susceptibilities must involve
full two-band operators represented by  $2\times 2$ matrices,
\begin{equation}
\begin{split}
j^\alpha_{ll' }(\bk)  &\equiv e\,  {\pz h\nd_{ll'} (\bk)\over\pz k_\alpha}\\
M_{ll'}(\bk) &\equiv   {ie \over 2c}  \bigg(  j_{ll' }^y (\bk)\,  {\pz\over\pz k_x} -j_{ll' }^x (\bk)\,  {\pz\over\pz k_y}\bigg)\quad,
\label{2b-curr}
\end{split}
\end{equation}
which yield the interband susceptibilities which can be conveniently expressed by \cite{SM},
\begin{align}
\chi_\cmc^\inter &=   \!\int \!{d^d k\over (2\pi)^d } \sum_{i=1}^2   f\big(\veki\big)\, F^\inter_i(\bk) \label{2b-susc}\nonumber\\
F^\inter_i(\bk)&= e \bigg[U\yd_\bk\, \bigg( {\pz\over\pz k_y}\,\big[M,j^x\big] -{\pz\over\pz k_x}\big[M,j^y\big]\bigg) 
U_\bk\bigg]\nd_{\!ii}\nonumber\\
\chi_\csr^\inter &= e \!\int \!{d^d k\over (2\pi)^d }\! \sum_{i=1}^2   f\big(\veki\big)\, 
\bigg[U_\bk\yd\, {\pz j^x(\bk)\over\pz k_x}\, U_\bk\bigg]_{\!ii}
\end{align}
The unitary matrix $U_\bk$ diagonalizes $h(\bk)$.  
We note that the operator $\pz\big[M,j^\alpha\big]/\pz k^\beta$ includes a derivative $\pz/\pz k_\alpha$ acting to the right on $U_\bk$.
This derivative captures the effect of SU(2) rotation of Bloch states inside the Fermi volume.

We now apply Eqs.~(\ref{csr-intra}) and (\ref{2b-susc})  to calculate the Hall coefficients of the following models.  
\begin{figure}[t]
	\includegraphics[width=0.48\textwidth]{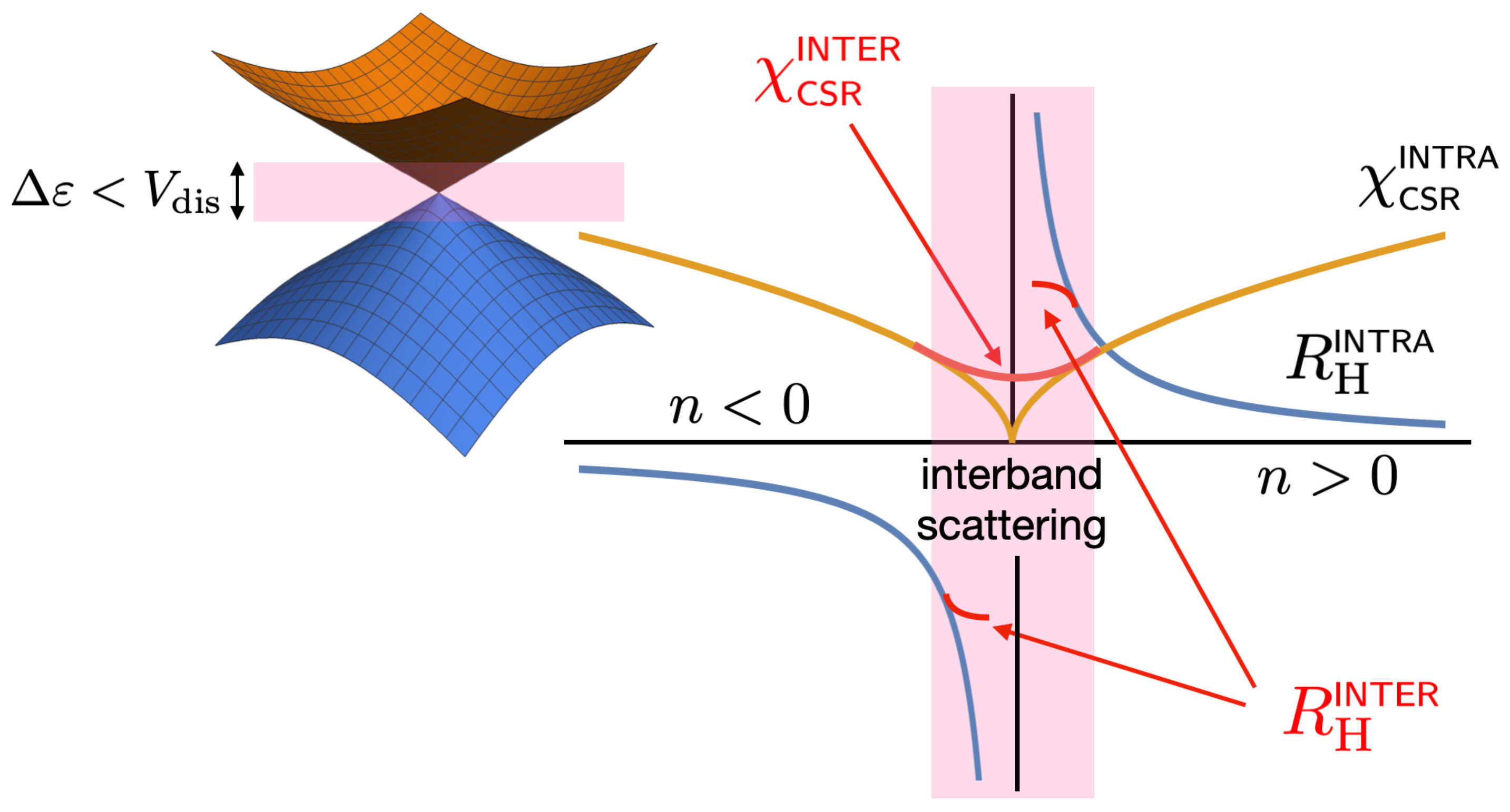}
	\caption{The  two dimensional Weyl cone, whose bands are depicted in the upper left. 
		The  intraband Hall coefficient  (online, blue) and conductivity sum rule
		(online, orange) are  plotted versus the density of carriers  $n$ as measured from the nodal filling. Pink (online) regions mark the low density interband dominated transport regime,
		where the interband gap is lower than the disorder potential $V_{\rm dis}$. In this regime, the conductivity sum rule $\chi^\inter_\csr$ does not vanish at the nodal density, and the Drude-like divergence of the Hall coefficient
		is cut off (see text). 
	}
	\label{fig:2dWeyl}
\end{figure}

{\em Weyl semimetals}.\textemdash  When the product of time reversal and inversion is not a symmetry of a system, the band structure may
exhibit Weyl points, where two bands intersect at the Fermi level.  Expansion of the semimetal band structure
near a linear point degeneracy results in  the $d$-dimensional  2-band Weyl Hamiltonian \cite{Comm-WeylSM}
\be
H\nd_0=v\nd_0\,\bk\cdot\bsigma
\ee
which yields the conical  dispersion $\ve\nd_\pm(\bk)=\pm v_0\, |\bk|$, see Fig.~\ref{fig:2dWeyl}. 
For $d=2$,  this could describe surface states of a  three dimensional topological insulator~\cite{TI1}, or a single
Dirac cone in graphene \cite{Graphene-CN}.  For $d=3$, this could describe one Weyl cone in a Weyl semimetal.

The density  (per cone) is $n= \sgn(n)\, k_{\rm F}^{d}/2d \pi^{d-1} $, where $\kF$ is the Fermi wavevector.  In the intraband transport regime,
\begin{equation}
\begin{split}
\chi_\cmc^\intra&=  {e^4 v_0^2 \over  c} { k_{\rm F}^{d-2}\over 2 d \pi^{d-1} } \, \sgn(n)\propto  \sgn(n)\, |n|^{(d-2)/d} \\
\chi_\csr^\intra&=  {v_0\over 2d \pi^{d-1} } k_F^{d-1}  \propto |n|^{(d-1)/d}\quad,
\label{Weyl-intra}
\end{split}
\end{equation}
which recovers the Drude relation $R^\intra_H =1/nec$.

For the interband regime \cite{SM} we find that,
\begin{equation}
\chi^\inter_\cmc (n)=\chi_\cmc^\intra(n)\quad , \quad   \chi^\inter_{\rm CSR}(n) \propto n^0.
\label{Weyl-2b}
\end{equation}

At low densities, the interband regime takes over when  $\Delta\ve < V\nd_{\rm dis}$, as depicted 
by pink (online) shaded areas in Fig.~\ref{fig:2dWeyl}. 
Since the sum rule in Eq.~(\ref{Weyl-2b}) does not vanish at the Weyl point  the Drude-like divergence of  the Hall coefficient is cut off at the Weyl point.

Unfortunately,  a quantitative calculation of $R^\inter_{\rm H}$ in this regime is not available, since the Fermi energy
is half the interband gap. This violates the ``good quasiparticles'' condition, and ${{\cal R}^{\rm corr}}$ cannot be neglected (as explained later).
Nevertheless,  since $\chi^\inter_\csr > 0$,  the  saturation of $R_{\rm H}^{\inter} <\infty$ at the Weyl point still holds.

{\em  Nodal-line  semimetal}.\textemdash
It is also possible for two bands to touch along a curve, as is the case in a
nodal line semimetal \cite{NodalSM1,NodalSM4}.  Such a state of affairs has reportedly been observed in the compound
ZrSiSe \cite{NLS_basov} as well as in optical lattices with ultracold fermions \cite{NLS_optical}.

We  consider a nodal  circle of radius $k_0$ in the $k_z$=0  plane, as depicted in Fig.~\ref{fig:NLSM}.  
The dispersions near the nodal line are expanded for low values of $\delta k_\perp=\sqrt{k_x^2+k_y^2}-k_0$ and $k_z $,
\begin{equation}
\ve_{\bk\pm}\simeq \pm v_0 \sqrt{\alpha^2\big(\delta k\nd_\perp\big)^{\!2} + k_z^2} 
\end{equation}
where $\alpha$ is a dimensionless anisotropy parameter.
Here we limit the calculation to the intraband regime at zero temperature, where $n = {k_0 \veF^2/4\pi\alpha v_0^2}$.   
By Eq. (\ref{csr-intra}),  the susceptibilities are
\begin{equation}
\chi_\cmc^\intra= {3 e^3 v^2_0 \alpha^2 n \over 4 k_0^2  c } \quad,\quad
\chi_\csr^\intra  = e^2 v_0  \left(  {\alpha^3 k_0 n \over 16 \pi} \right)^{\!\!1/2},
\end{equation}
which yields  an unusual density dependence of the Hall coefficient,
\begin{equation}
R^\intra_{\rm H} =  {12 \pi \over \alpha k_0^3 e c} ~\sgn(n)
\end{equation}
The nodal line semimetal exhibits a density independent Hall coefficient with an abrupt sign reversal, at zero temperature and disorder.  { The  suppression of  $\chi_\cmc^\intra$  at large radii  can be intuitively attributed to the
	near cancellation of the inner (hole-like) and outer (electron-like) sides of the toroidal Fermi surface.}

The density dependences of Weyl and nodal line semimetals are summarized in Table \ref{table1}. 

\begin{table}[t!]
	\begin{center}
		\begin{tabular}{|c|c|c|c|c|}
			\hline
			Model  & $\chi_\csr^\intra$  & $R_{\rm H}^\intra$  & $\chi_\csr^\inter$  & $ R_{\rm H}^\inter$   \\
			\hline
			2d Weyl   & $  |n|^{1/2}$   & $    1/n  $ &const &$    \le {\rm const}$   \\
			3d Weyl   & $|n|^{2/3}$   & $  1/n  $ & const & $  \le  |n|^{1/3} $   \\
			nodal line sm  & $ |n|^{1/2}$   & $ \sgn(n) $ & $~$ & $~ $   \\
			\hline
		\end{tabular}
		\caption{Nodal line semimetal and Weyl  semimetals in 2 and 3 dimensions. The density dependence of the conductivity sum rules and Hall coefficients 
			are given for the intraband and interband transport regimes.}\label{table}
		\label{table1}
	\end{center}
\end{table}

\medskip
{\em Semiconductor bands}.\textemdash   with an inversion-asymmetric zinc blende structure, e.g. GaAs and CdTe, are subjected to spin orbit interactions described by the Kane and Luttinger models \cite{KaneModel, Luttinger, Winkler}.  
They  share with semimetals the small interband gaps near the Fermi energy. We study two models:
(i) The (heavy) hole bands in a two dimensional quantum well (2dH) \cite{Culcer}:
\begin{equation}
h^\tdh(\bk) = {k^2 \over 2m}\mathbb{I} \ \pm\  \beta \left[k_y(k_y^2-3k_x^2)\,\sigma^x + k_x(k_x^2-3k_y^2)\,\sigma^y \right]
\label{2dH}
\end{equation}
where the Rashba parameter $\beta$ depends on the perpendicular electric field \cite{Culcer}. The bands 
$\ve^\tdh_{\bk\pm}  = {k^2/2m} \pm \beta k^3$ are rotationally symmetric, and split by $\beta$. 

(ii) The conduction band  in a cubic crystal,   with spin orbit interaction splitting expanded up to third order in $k$~\cite{Winkler},
\begin{align}
h^\tdc(\bk) &= {k^2 \over 2m}\mathbb{I}  \pm  \beta\,  \bh(\bk) \cdot \bsigma \label{3dC}\\
\bh(\bk) &=\big(k_y^2 - k_z^2\big) k_x\xhat +\big (k_z^2 - k_x^2\big) k_y \yhat + \big(k_x^2 - k_y^2\big) k_z\zhat \nonumber
\end{align}
the dispersions $\ve^\tdc_{\bk\pm} =  {k^2/ 2m}\pm \beta |\bh_\bk|$, have cubic symmetry.

We find that for both models, Eq.~(\ref{2dH}) and \ref{3dC}),  the susceptibilities and Hall coefficients are corrected by  terms of order order $\beta^2$:
\begin{equation}
\chi_\csr  = {e^2 \over m}\left(n +\beta^2  \Delta \chi_\csr\right) , \quad  R_{\rm H} = {1+   \beta^2 \CK (n)\over nec}\quad,
\label{DRH}
\end{equation}
The results for the corrections of both Eq.~(\ref{2dH}) and (\ref{3dC}) are listed in Table \ref{table2}.
The density dependence and sign of the intraband corrections for  the heavy holes model (\ref{2dH}) are consistent  with Ref.~ \cite{Culcer}.
Our new results for the interband regime \cite{SM} show that while $\chi^\inter_\cmc=\chi^\intra_\cmc$, the sum rule is different, since it acquires no order $\beta^2$ corrections, i.e. $ \chi^\inter_\csr={e^2 n\over m} $.  
As a result, we obtain that $\CK^\inter = -\CK^\intra$, that is to say, the spin-orbit correction to the Drude Hall coefficient  reverses sign as disorder increases between the   intraband and   interband  scattering regimes.

\begin{figure}[t]
	\includegraphics[width=0.3\textwidth]{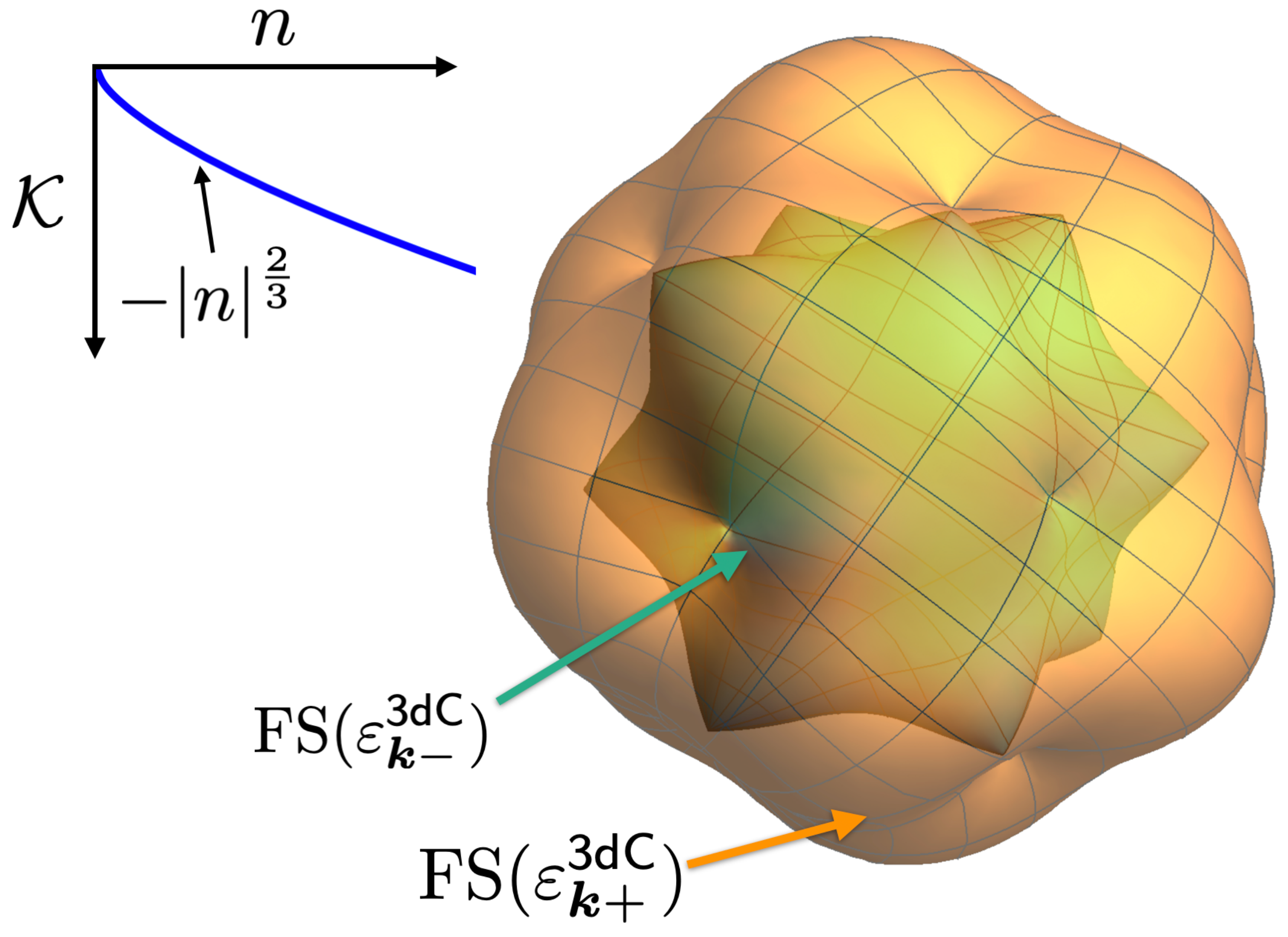}
	\caption{Spin-orbit split Fermi surfaces  (FS) of conduction electrons described by the Hamiltonian Eq.~(\ref{3dC}). Top left:  Density dependence of the non-Drude correction $\CK$, Eq.~(\ref{DRH}). }
	\label{fig:3dC}
\end{figure}
For $h^\tdc(\bk) $,  the spin-orbit correction $\Delta \chi^\tdc_\csr$ is of order $- n^{5/3}$ due to the $k^3$ scaling of  $\bh(\bk)$.  The 
interband susceptibility $\chi^\inter_\cmc$ is not equal in magnitude to $ \chi^\intra_\cmc  $, which appears to be due to non-spherical symmetry of the bandstructure, as shown in Fig.~\ref{fig:3dC}.       
\begin{table}[t!]
	\begin{center}
		\begin{tabular}{|c|c|c| c|}
			\hline
			Model  & $\Delta \chi_\csr^\intra/m^2$  & $\CK^\intra/m^2$  &  $\CK^\inter/ m^2$   \\
			\hline
			2dH &  $ - 18 \pi   n^2   $  & $18\pi   n $   &  $-18\pi   n    $   \\
			3dC  & $- 8.0(1)\, n^{5/3}  $   & $ - 17.5(1)\,n^{2/3}  $ &   $ -23.0(1)\,n^{2/3 }  $   \\
			\hline
		\end{tabular}
		\caption{Spin-orbit corrections to  the sum rule and Hall coefficient factor for the two dimensional hole bands Eq.~(\ref{2dH}), and three dimensional conduction bands, Eq. (\ref{3dC}).
			Results for the intraband and interband transport regimes are displayed. The conductivity sum rule receives no order $\beta^2$ correction in the  interband  regime.} \label{table2}
	\end{center}
\end{table}

{ 
	{\em Estimation of the correction $\CR$.}\textemdash
	We now prove that $\CR$,  of Eq.~(\ref{Rcorr}),  vanishes as  (at least) two powers of the disorder potential over the Fermi energy. Explicit instructions to calculate the moments, recurrents, Krylov bases and magnetization matrix elements \cite{EMT} are reviewed in \cite{SM}.
	Let us first consider the intraband  scattering regime where $V_{\rm dis} \ll \Delta\ve \ll \veF$:

	The intraband currents  commute with the clean Hamiltonian $[\CH_0,\tj^x]=0$. Hence the high order Krylov operators are produced by commuting the current with at least one power of the disorder potential.  The magnetization matrix elements between normalized Krylov bases should therefore scale as,
	\be
	(1-\delta_{i,0}\delta_{j,0} )| \CM''_{2i,2j} |  \propto    {\ccmc \over \ccsr } \left(  {V_{\rm dis} \over \veF}   + \CO\left( {V_{\rm dis} \over \veF} \right)^2 \right) 
	\label{CM-intra}
	\ee
	For similar reasons, the first two conductivity moments scale as,
	\be
	\begin{split}
		\mu_2 &  \propto    \ccsr\,  V^2_{\rm dis}    \\
		\mu_4 &      \propto   \ccsr\,\left( \veF^2 V^2_{\rm dis}   +  \CO ( V^4_{\rm dis} )\right),
		\label{mu2i}
	\end{split}
\end{equation}
Transforming moments to recurrents  (see \cite{SM}) yields the ratio,
\begin{equation}
{\Delta_1\over \Delta_2} =\left(  {\mu^2_2 \over   \ccsr\, \mu\nd_4-\mu^2_2 }\right)^{1\over 2}  \propto  {V_{\rm dis} \over \veF } .
\label{Delta12}
\end{equation}

Combining (\ref{CM-intra}) and (\ref{Delta12})  in (\ref{Rcorr}), we obtain an overall multiplicative factor,
\begin{equation}
|\CR|  \propto   { |\ccmc | \over \chi^2_\csr } \left(  {V_{\rm dis}^2 \over \veF^2} \right).
\label{Rcorr-est}
\end{equation}
In the metallic phase, $R_{\rm H}, \ccmc/(\ccsr)^2<\infty$,  and hence the infinite sum in $\CR$ must converge. Therefore the coefficient of proportionality in (\ref{Rcorr-est}) must be finite.

For the interband regime, we use Eqs.~(\ref{Hk},\ref{2b-curr})) to obtain $ \mu_2=||~[\CH , j^x]~||^2 \propto   \big(V^2_{\rm dis} +(\Delta\ve)^2\big)  \ccsr $.
We also assume $(\Delta\ve)^2  \le V^2_{\rm dis}$. Thus, we can appeal again to 
Eqs. (\ref{CM-intra},\ref{Delta12}) by simply replacing $V^2_{\rm dis} \to V^2_{\rm dis} + (\Delta\ve)^2  \le 2 V^2_{\rm dis} $.
This recovers the same proportionality if  Eq.~(\ref{Rcorr-est}) as applicable also to the interband regime, where we use $ \chi_\cmc^\inter/ (\chi_\csr^\inter )^2  $ to
compute the  Hall coefficient  \cite{Comm-inter-intra}.
Thus  $\CR$ can be neglected relative to ratio of corresponding susceptibilities as long as $  {V_{\rm dis}^2 \ll  \veF^2} $  in both intraband and interband regimes.

{\em Summary}.\textemdash  Eq.~(\ref{RH-gen}) provides insight into  deviations from Drude's relation in semimetals. Our calculations demonstrate  the effects of non-spherical and multiple Fermi surfaces,
and interband scattering. These effects should be considered when comparing the ``Hall number'' ($R_{\rm H}^{-1}$) to the Fermi volume, as determined   by  e.g. angular resolved photoemmision~\cite{TI2}, 
and magneto-transport oscillations \cite{SdH,FS-MacKenzie}.   {  For time reversal invariant Weyl  semimetals,
	topologically protected surface states have been shown \cite{Piet} to contribute substantially to the {\em longitudinal } conductivity in small samples. Future investigations of the finite size corrections to the Hall coefficient due to surface Fermi arcs states would be interesting.  }
For graphene, we  propose to split the Dirac cones by an in plane magnetic field. The Hall coefficient should vanish  between gate voltages $ V_{\rm gate}= \pm g \mu_B B/e$, 
which may enable measurements of  the compressibility  at low densities.

{\em Acknowledgement --}   
We acknowledge support from the US-Israel Binational
Science Foundation Grant No. 2016168 and the Israel
Science Foundation Grant No. 2021367. This research was
also supported in part by the ICTS/topmatter2019/12
Program and the program at KITP Santa Barbara, funded
by the National Science Foundation under Grant No. NSF
PHY-1748958.

\bibliographystyle{unsrt}
\bibliography{refs.bib}

\begin{thebibliography}{10}
\bibitem{Comm-ISL}
{ All perturbations to the band Hamiltonian, including those which give rise to
  anisotropic $\bk$-dependent scattering rates, contribute only to $R^{\rm
  corr}$, which by Eq.~(\ref{Rcorr-est}) vanishes for weak disorder.}
\bibitem{Comm-WeylSM}
{A tilted Weyl cone model includes an additional $ \bu\cdot \bk\mathbb{I} $,
  which for large $|\bu|$ defines a type-II Weyl semimetal \cite{Soluyanov},
  with large electron and hole pockets at the Weyl point filling. For
  $0<|\bu|<1$, $R_{\rm H}$ is evaluated in \cite{SM}.}
\bibitem{Comm-inter-intra}
{ It is not advisable to use $\chi_\cmc^\inter,\chi_\csr^\inter$ in the
  intraband scattering regime. The commutator of the Hamiltonian with the
  interband current $ j^x_{12 }(\bk) $ produces a correction $\CR \propto
  (\Delta\ve)^2$, which does not vanish in the zero disorder limit. The use of
  the projected intraband current ensures that the correction vanishes at weak
  disorder. }.
\end{thebibliography}

\clearpage
\begin{widetext}
\begin{center}
\textbf{\large  Supplementary Material for: Hall coefficient of  semimetals}
\end{center}

\section{Current cross susceptibilities}
Here we derive the expressions leading to Eqs. (6) and (8) in the main text.
The general expression of susceptibility between two second quantized operators $\hj^\alpha$ and $\hA$ is \cite{EMT},
\begin{equation}
	(\hj^\alpha,\hA) = {1\over Z} \sum_{nm} {e^{-\beta E_n} - e^{\-\beta E_m} \over E_m-E_n } \langle n|  \hj^\alpha |m\rangle  \langle m| \hA |n\rangle ,
\end{equation}
where $Z= \Tr e^{-\beta\cH}$, $\beta$ is the inverse temperature, and $\cH$ is the full Hamiltonian with  spectrum $\{E_n,|n\rangle\}$. 
This current susceptibility can be written as an expectation value using the polarization operator 
\begin{equation}
	(\hj^\alpha,\hA)= {1\over Z} \sum_{n }  e^{-\beta E_n}  \Im \langle n|\left[ \hP^\alpha,\hA\right] |n\rangle \equiv   \Im \langle \left[ \hP^\alpha,\hA\right] \rangle ,
	\label{expect}
\end{equation}
where $\hP^\alpha$ is defined by Ehrenfest relation,
\begin{equation}
	\left[ \cH,\hP^\alpha \right] = i \hj^\alpha
\end{equation}

For band electrons, translationally invariant single particle operators are represented by the bilinear forms,
\begin{eqnarray}
	&&{\hat A}= \sum_{\bk, ll'} c^\dagger_{\bk,l} A_{\bk  ll'} c_{\bk,l'} ,\quad \hj= \sum_{\bk, ll'} c^\dagger_{\bk,l} j^\alpha_{\bk,ll'} c_{\bk,l'} , \no\\
	&& \cH_0 = \sum_{\bk, ll'} c^\dagger_{\bk,l} h_{\bk,ll'} c_{\bk,l'} ,\quad \hP^\alpha = i \sum_{\bk, l} c^\dagger_{\bk,l}  \partial_{k_\alpha}  c_{\bk,l} ,
\end{eqnarray}
where $l,l'$  are  band indices. The susceptibilities are given by the integrals,
\begin{equation}
({\hat j}^\alpha, {\hat A}) =  \lim_{{\bf q}\to 0} \sum_{ij}  {f(\ve_{\bk i})  - f(\ve_{\bk+{\bf q} j} )\over \ve_{\bk +{\bf q} j} -\ve_{\bk  i}  }~ j_{\bk i,\bk+{\bf q} j}^\alpha ~   A_{\bk+{\bf q} j,\bk i}~ .
\label{full}
\end{equation}

The intraband currents are diagonal in the eigenstates with dispersion $\veki= (U^\dagger_\bk h_\bk U_\bk )_{ii}$. Thus, in Eq.~(\ref{full})
only terms with $i=j$ survive, and  $  {f(\ve_{\bk + {\bf q} i} ) - f(\ve_{\bk  i}) \over \ve_{\bk+{\bf q} i} -\ve_{\bk  i}  } \to \partial_\ve f(\veki) $ leading to Eqs. (6) in the main text.
For the full interband currents, the susceptibilities are more conveniently expressed using Eq.~(\ref{expect}):
\begin{equation}
	(\hj^\alpha, \hA) = \sum_{\bk,i}   f(\veki)    \left(U_\bk^\dagger \left( \partial_{k_\alpha} A_{\bk }\right)  U_{\bk } \right)_{ii} ,
\end{equation}
which yield Eqs. (8) in the main text.
\section{Weyl semimetals}
A general Weyl Hamiltonian is
\be
\cH= v_0 \left( \sum_{i=1}^d a_i k_i \sigma^i + u k_z \mathbb{I}\right)
\ee
where $u\ne 0$ describes a tilted Weyl Hamiltonian. For $u=0$, it is possible to substitute $k_i'= a_i k_i$ and map the Brillouin zone integrals to those of a spherical band. This for $u=0$ we restrict ourselves to $a_i=1$. 

\subsection{Intraband regime, $u=0$}

Near the Weyl point, the dispersion is spherically symmetric
\be
\vek=\pm v_0 |\bk|
\ee
Using Eq.~(2) (main text) noting that $\ve'(k_F) =v_0$,  we  obtain  the intraband susceptibilities in two dimensions as
\bea
n={k_F^2 \over 4\pi},\quad
\chi_\cmc^\intra =   {e^3 v_0^2 \over 4\pi c}~  \sgn(\epsilon_F),\quad
\chi_\csr^\intra =  {e^2 v_0  \over 4\pi }~ k_F     
\eea
and in three dimensions
\bea
n= {k_F^3 \over 6\pi^2},\quad
\chi_\cmc^\intra =   {e^3 v_0^2 k_F \over 6\pi^2 c}~  \sgn(\epsilon_F) ,\quad
\chi_\csr^\intra =   {e^2 v_0  \over 6\pi^2 }~ k^2_F     
\label{untilted}
\eea
Both two and three dimensions recover Drude's relation for $R_{\rm H}$, although they differ in the density dependences of the individual susceptibilities.
\subsection{Tilted 3d Weyl cone $0<u<1$}
Using polar coordinates  $\bk=(k,\theta,\phi)$, the dispersion and radial velocities are
\be
\ve(\bk)= v_0 k (1+u\cos \theta ), 
\ee
Defining the equator Fermi wavevector by $k_F= {\ve_F\over v_0}$, we obtain,
\bea
\chi_\cmc^\intra &=& {e^3 v_0^2 \over 12 \pi^2 c} k_F  \int_{-1}^{1}d\cos\theta~ (1+u\cos\theta)\sin^2\theta =  {e^3 v_0^2  \over 6\pi^2 c}k_F~  \sgn(\epsilon_F) ,\quad\nonumber\\
\chi_\csr^\intra &=&  {e^2 v_0  \over 12\pi^2 }~ k^2_F   \int_{-1}^{1} \! d\cos\theta~{\sin^2\theta\over   1+u\cos\theta }   = {e^2 v_0  \over 6\pi^2 }~ k^2_F   \left( {u +(u^2-1) \mbox{\rm arctanh}(u)\over u^3} \right)
\eea
which yields a Hall coefficient at small $u$:
\be
R^\intra_{\rm H} \simeq {1\over nec}\left( 1- {2\over 5}u^2 + \cO(u^4)\right)
\ee
\subsection{Interband regime, $u=0$}
The two band  current is wave-vector independent, 
$j^\alpha = ev_0 \sigma_\alpha $ and therefore the integrands in $\ccsr, \ccmc$ acquire contributions only from the band  bottom,  while the integrand of $\ccmc$  also depends on the wavefunction
rotation matrix $(\partial_{k_\alpha}U_\bk) U_\bk^\dagger$ around the Weyl singularity:
\bea
U^\dagger_\bk {e \over c}\partial_\alpha [M,j^\beta] U_\bk &=&  \epsilon_{\alpha\beta} {e^3v_0^2 \over  2c} [\sigma_\alpha,\sigma_\beta] \partial_\beta (\partial_\alpha U_\bk) U_\bk^\dagger  
={e^3v_0^2\over  c}   \pi \delta^2(\bk)  \nonumber\\
\chi_\cmc^\inter&=&(\ccmc)_0+{e^3v_0^2\over  c}  \int {d^d k\over (2\pi)^d }(f(v_0 k) + f(-v_0 k))  \delta^2 (\bk)   = \chi_\cmc^\intra  \nonumber\\
\chi_\csr^\inter&=&{\rm const}
\eea
\section{Nodal line semimetal}
The dispersions for the nodal-line semimetal in three dimensions are expanded near the nodal line for small values of $\delta k_\perp=\sqrt{k_x^2+k_y^2}-k_0$ and $k_z $
\begin{equation}
	\ve_{\bk\pm}\simeq \pm v_0 \sqrt{\alpha^2\big(\delta k\nd_\perp\big)^{\!2} + k_z^2} 
\end{equation}
which corresponds to a nodal circle of radius $k_0$ in the $k_z=0$ plane. 
$\alpha$ is a dimensionless anisotropy parameter, given by $\alpha={k_0 \over \sqrt{m^2+k_0^2}}$.

The density is related to the Fermi energy by
\begin{equation}
	n = {k_0 \veF^2 \over 4\pi\alpha v_0^2}.
\end{equation}
The intraband velocities and their derivatives are given by
\begin{eqnarray}
	v^x(\bk) &=& {\partial \ek \over \partial k_x} = {v_0^2\alpha^2 \over k_0 \ek} k_x\delta k_\perp ,\quad
	v^y(\bk) = {\partial \ek \over \partial k_y} = {v_0^2\alpha^2 \over k_0 \ek} k_y\delta k_\perp \no\\
	{\partial v^x(\bk) \over \partial k_x}
	&\approx& {v_0^2\alpha^2 \over k_0 \ek}  \lt( \delta k_\perp + {k_x^2\over k_0} \rt)  ,\quad
	{\partial v^y(\bk) \over \partial k_y } \approx  {v_0^2\alpha^2 \over k_0 \ek} \lt( \delta k_\perp + {k_y^2\over k_0} \rt),\quad
	{\partial v^x(\bk) \over \partial k_y}
	\approx  {v_0^2\alpha^2 \over k_0 \ek} {k_x k_y \over k_0} .
\end{eqnarray}
Now using $k_x^2+k_y^2\approx k_0^2 + 2k_0\delta k_\perp$, the conductivity sum rule at zero temperature is calculated to be,
\begin{eqnarray}
	\chi_\csr^\intra= {e^2\over (2\pi)^3}\int dk_z \int dk_\perp d\theta\ k_\perp \ {\alpha^4 \delta k_\perp^2 \over \ek^2}\ {1\over 2}\lt(1+2{\delta k_\perp \over k_0}\rt)\ \delta (\ek-\veF)  
	=e^2{\alpha k_0 \veF \over 8 \pi}
\end{eqnarray}
The mean Fermi surface curvature is given by, 
\begin{eqnarray}
	F(\bk) = {1\over 2} \lt( [v^x(\bk)]^2{\partial v^y(\bk) \over \partial k_y} + [v^y(\bk)]^2{\partial v^x(\bk) \over \partial k_x} - 2v^x(\bk) v^y(\bk) {\partial v^x(\bk) \over \partial k_y} \rt) 
	\approx  {v_0^6\alpha^6  \over 2 \ek^3 k_0^3}\ \delta k_\perp^3 (k_0^2+2k_0\delta k_\perp)
\end{eqnarray}
Using this, the current-magnetization-current susceptibility at zero temperature is given by,
\begin{eqnarray}
	\chi_\cmc^\intra ={e^3\over c}{1\over (2\pi)^3} {v_0^6\alpha^6  \over k_0^2} \int dk_z \int dk_\perp d\theta\ k_\perp {\delta k_\perp^4\over \ek^3}\ \delta (\ek-\veF) 
	= {3e^3\alpha \veF^2 \over 16c\pi k_0}
\end{eqnarray}
Therefore we find,
\begin{equation}
	R^\intra_{\rm H} =  {\chi_\cmc^\intra \over \lt(\chi_\csr^\intra\rt)^2}
	= {12 \pi \over \alpha k_0^3ec} ~\sgn(n) .
\end{equation}
\section{Heavy holes model}
The full two band model of  spin-orbit split heavy holes band \cite{Culcer} bands is
\bea
H_{\bk}^{\rm 2dH} = {k^2 \over 2m} + h^x_{\bf k}\sigma_x + h^y_{\bf k}\sigma_y 
= {k^2 \over 2m} + \beta k^3 \hat\tbphi_\bk\cdot\vsigma
\label{2dHH}
\eea
where   $\beta$ is the Rashba coefficient, and $ \hat\tbphi_\bk$ is a unit vector in the direction $\tphi_\bk   = 3\phi_\bk+{\pi\over 2}$.

The spectrum  is,
\begin{equation}
	\ek^{\pm}= {k^2 \over 2m}\pm \beta k^3
\end{equation}
which  yields two Fermi circles with radii difference $k_{F_+}-k_{F_-}=\Delta k_F =- 2m\beta k_F^2$, where $k_F=(k_F^+ + k_F^-)/2$. The two radial velocities are
\be
\partial_k \epsilon_i  = { k\over m}\pm 3 \beta k^2
\ee
\subsection{Intraband regime} 
For the intraband susceptibilities of two concentric spherical fermi surfaces, we can use the formula 
$ R^\intra_{\rm H} = {\sum_i \ccmc(i) \over (\sum_i \ccsr(i))^2}    $,
where for each band separately $\ccmc(i), \ccsr(i)$ are given by Eq. (2) of the main text.
An thus, up to order $\beta^2$, we obtain the following quantities, for $n={k_F^2\over 2\pi},$ 
\bea
\chi_\cmc^\intra = 
{ e^3\over m^2 c} (n- 18 \pi m^2\beta^2 n^2),\quad
\chi_\csr^\intra  = { e^2 \over m}   (n - 18 \pi m^2\beta^2 n^2),\quad
R^\intra_{\rm H}= {1\over nec} \left( 1 + 18\pi m^2\beta^2 n \right)
\label{HHintra}
\eea
\subsection{Interband regime}
The unitary transformation which diagonalizes the Eq.~(\ref{2dHH})   is  
\be
U_\bk =  e^{ -{i\over 2} \tphi_\bk  \sigma^z } e^{ - {i\over 4} \pi \sigma^y} 
\label{U}
\ee
The velocity matrices are given by,
\begin{eqnarray}
	v^x_{\bf k} &=& \partial_{k_x} H_\bk = {k_x\over m} - 6\beta k_x k_y \sigma^x + 3\beta (k_x^2-k_y^2)\sigma^y , \no\\
	v^y_{\bf k} &=& \partial_{k_y} H_\bk = {k_y\over m} + 3 \beta (k_y^2-  k_x^2) \sigma^x - 6\beta k_x k_y  \sigma^y .
	\label{V}
\end{eqnarray}

The sum rule is given by rotating the operator  ${\partial\over\partial k_x}v_\bk^x $ onto the $\sigma^z$ axis using  Eq. (\ref{U}). 
\begin{eqnarray}
	\chi_\csr^\inter=   e^2 \sum_{\bk,i=\pm} f_i \left(U_\bk^\dagger \left({\partial\over \partial k_x} v_\bk^x \right)U_\bk\right)_{ii} 
	= {e^2 \over m} \sum_{\bk,i=\pm} f_i +  6 \beta  \sum_{\bk} (f_+-f_-)  k  \cos(2\phi_\bk)  
	= {n e^2 \over m}
	\label{sumrule}
\end{eqnarray}
where the second term vanishes  by circular symmetry of the band structure  and $\int_0^{2\pi} \! d\phi_\bk\cos(2\phi_\bk)=0$.

The magnetization matrix  operator of Eq.~(6) of the main text is,
\begin{equation}
	M_{ss'} (\bk) = {ie\over 2c} \left( v_\bk^y {\partial \over \partial k_x} -v_\bk^x {\partial \over \partial k_y} \right)_{ss'}
\end{equation}
Commuting $M$ with the velocities yields an anti-hermitian operator
\begin{eqnarray}
	\left[M,v_\bk^y\right] &=& { ie\over 4c}  \left( v_\bk^y    \partial_{k_x} v_\bk^y-v_\bk^x   \partial_{k_y} v_\bk^y  -\left[v_\bk^x,v_\bk^y \right]   \partial_{k_y} + {\rm h.c.} \right) \nonumber\\
	\left[v^x_\bk,v^y_\bk\right] &=& i\beta^2 18 \left( 4  k_x^2 k_y^2 +   (k_x^2 -k_y^2)^2 \right)\sigma^z =i \beta^2 18 k^4 \sigma^z.
	\eea
	The operator in $\ccmc$ is 
	\begin{eqnarray}
		{\cal M}_\bk &=& ie\left( \partial_{ k_x} (e\left[M,v_\bk^y\right]) - \partial_{ k_y}(e\left[M,v_\bk^x\right])\right) \nonumber\\
		&& = {e^3\over  c} \left( {1\over m^2} - 72\beta^2 k^2 +  \beta^2 k^2 i \sigma^z \left(   36 ~\bk \times\nabla_\bk  +9 k^2 \nabla_\bk \times\nabla_\bk\right)\right)
	\end{eqnarray}
	The transformation of the last term to the Hamiltonian eigenbasis is given by
	\begin{equation}
		i  U_\bk^\dagger    \sigma^z \left(   36 ~\bk \times\nabla_\bk  +9 k^2 \nabla_\bk \times\nabla_\bk\right) U_\bk =   {3\over 2}  \left(   36 ~ \bk \times \nabla_\bk \phi_\bk + 18 \pi k^2 \delta^2(\bk)\right)=  54\beta^2 k^2 
	\end{equation}
	where the $\delta^2(\bk)$ does not contribute because of the prefactor of $k^4$. Hence
	\begin{equation}
		(U_\bk^\dagger {\cal M} U_\bk)_{ii'} =  {e^3\over m^2c} \left(  1  - 18\beta^2 k^2\right) \delta_{ii'}
	\end{equation}
	Thus we obtain the Hall coefficient to order $\beta^2$,
	\begin{equation}
		R^\inter_{\rm H}  = {1\over nec} \left( 1- 18m^2\beta^2 {1\over N} \sum_{\bk,i}  f(\ek^i) k^2\right)={1\over nec} \left( 1- 18\pi m^2\beta^2 n \right)
	\end{equation}
	where we note that the sign of the correction is opposite to that of $R^\intra_{\rm H} $ in Eq.~(\ref{HHintra}).
	
	\section{3D Conduction band}
	For the conduction band~\cite{Winkler} in three dimension, the Hamiltonian is given by,
	\be
	H_\bk^{\rm 3dC} = {k^2\over 2m} + \beta  \left[ (k_y^2 - k_z^2) k_x \sigma^x +(k_z^2 - k_x^2) k_y \sigma^y+(k_x^2 - k_y^2) k_z \sigma^z \right]
	\ee
	where the spectrum is ,
	\begin{equation}
		\ek_\pm =  {k^2\over 2m} \pm \beta \sqrt{(k_y^2-k_z^2)^2k_x^2 + (k_z^2-k_x^2)^2k_y^2 + (k_x^2-k_y^2)^2k_z^2}
	\end{equation}
	and the unitary matrix which diagonalizes $h^{\rm 3dC}$ is 
	\bea
	U_\bk&=&  e^{-{i\over 2}\tphi_\bk \sigma^z}  e^{-{i\over 2}\ttheta_\bk \sigma^y}.\nonumber\\
	\tan\ttheta_\bk &=& {\sqrt{ k_x^2(k_y^2-k_z^2)^2 + k_y^2(k_z^2-k_x^2)^2 } 
		\over k_z(k_x^2-k_y^2)}   ,\quad
	\tan\tphi_\bk =  { k_y(k_z^2-k_x^2) \over k_x(k_y^2-k_z^2)}.  \eea
	
	\subsection{Intraband regime}

	We numerically evaluate the sum rule and the numerator of the magnetization, which behave as,
	\begin{eqnarray}
		\chi_\csr^\intra&=& {ne^2\over m} \lt(1-8.0(1)m^2\beta^2 n^{2/3}\rt) ,\quad
		\chi_\cmc^\intra =  {ne^3\over m^2c} \lt(1-33.5(1)m^2\beta^2 n^{2/3}\rt).
	\end{eqnarray}
	Therefore, the intraband Hall resistivity is given by,
	\begin{eqnarray}
		R_{\rm H}^\intra&=& {1\over nec}  \left( 1-17.5(1) m^2\beta^2 n^{2/3} \right).
	\end{eqnarray}
	\subsection{Interband regime}
	The velocities are
	\begin{eqnarray}
		v^x_\bk &=& {k_x\over m} +\beta   (k_y^2   - k_z^2)  \sigma^x  -2\beta k_x k_y\sigma^y +2\beta k_x k_z \sigma^z \nonumber\\
		v^y_\bk &=& {k_y\over m} +2 \beta   k_y k_x  \sigma^x   +  \beta (k_z^2-k_x^2) \sigma^y - 2\beta k_y k_z \sigma^z
	\end{eqnarray}

	The sum rule is 
	\begin{eqnarray}
		\chi_\csr^\inter &=& e^2 \sum_{\bk,i=\pm} f(\ek^i) \left(U_\bk^\dagger \left({\partial\over \partial k_x} v_\bk^x \right)U_\bk\right)_{ii}\nonumber\\ 
		&=& {ne^2 \over m} + 2 e^2\beta \sum_{\bk} \lt(f_+ -f_- \rt) k \left(\cos\theta_\bk\cos\ttheta_\bk   -\sin\theta_\bk\sin\phi_\bk\sin\ttheta_\bk\sin\tphi_\bk  \right) 
		= {ne^2\over m} 
		\label{SR3D}
	\end{eqnarray}
	The order $\beta$ term   vanishes under angular integration by symmetry.
	
	The commutator of the magnetization with the currents is, 
	\bea
	e\left[M,v_\bk^y\right] &=&  { ie^2\over 4c}  \left( v_\bk^y    \partial_{k_x} v_\bk^y-v_\bk^x   \partial_{k_y} v_\bk^y  -\left[v_\bk^x,v_\bk^y \right]   \partial_{k_y} + {\rm h.c.} \right) ,\quad
	\left[v^x_\bk,v^y_\bk\right]  =   2i\beta^2  {\cal C}_\bk , \nonumber\\
	{\cal C}_\bk &=&  k_x k_z (4k_y^2 + 2k_x^2 -2k_z^2)\sigma^x +  k_y k_z (4k_x^2 + 2k_y^2 -2k_z^2)\sigma^y+ (3k_x^2 k_y^2 + k_z^2 (k_x^2 + k_y^2 -k_z^2 )) \sigma^z.
	\eea
	Thus, the  operator in $\ccmc$ is 
	\begin{eqnarray}
		U_\bk^\dagger{\cal M}U_\bk &=& U_\bk^\dagger \lt\{ i e \left[{\partial\over \partial k_x} ,  e\left[M,v_\bk^y\right]\right] - ie\left[ {\partial\over \partial k_y}, e\left[ M,v_\bk^x\right] \right]\rt\}U_\bk\no\\ 
		&=& {e^3\over c}\left({1\over m^2} +{ 2\beta\over m} U_\bk^\dagger (k_x \sigma^x - k_y \sigma^y) U_\bk - 4\beta^2k^2 +\beta^2  U_\bk^\dagger {\cal R}_\bk  U_\bk \right),\nonumber\\
		U_\bk^\dagger {\cal R}_\bk  U_\bk  &=& \left( {\partial{\cal C}_\bk \over \partial k_x} {i\partial U_\bk\over \partial k_y}  - {\partial{\cal C}_\bk \over \partial k_y} {i\partial U_\bk \over \partial k_x}     +
		{\cal C}_\bk     \left({ \partial \over \partial k_x} {i\partial U_\bk \over \partial k_y}- { \partial \over \partial k_y}  { i\partial U_\bk\over \partial k_x}\right) \right) U_\bk^\dagger
		\label{UMU}
	\end{eqnarray}
	The order $\beta$ term in Eq.~\ref{UMU} vanishes upon integration.
	We define,
	\begin{eqnarray}
		{\partial{\cal C}_\bk \over \partial_{k_x}} &=& k_z(4k_y^2+6k_x^2-2k_z^2)\sigma^x + 8k_x  k_y k_z \sigma^y+k_x(6k_y^2+2k_z^2) \sigma^z \equiv \mbox{\boldmath{$A_x$}}\cdot \vsigma \no\\
		{\partial{\cal C}_\bk \over \partial_{k_y}} &=& 8k_y k_x k_z \sigma^x+  k_z(4k_x^2+6k_y^2-2k_z^2)\sigma^y +k_y(6k_x^2+2k_z^2) \sigma^z \equiv \mbox{\boldmath{$A_y$}}\cdot \vsigma
	\end{eqnarray}
	\begin{eqnarray}
		U_\bk &=& e^{-{i\over 2}\tphi_\bk \sigma^z}  e^{-{i\over 2}\ttheta_\bk \sigma^y}
		\equiv U_{\tilde\phi_\bk} U_{\tilde\theta_\bk} \no\\ \no\\
		i{\partial U_\bk \over \partial k_\alpha}U_\bk^\dagger &=& {1\over 2} {\partial \tilde\phi_\bk \over \partial k_\alpha}\sigma^z U_\bk + {1\over 2}{\partial \tilde\theta_\bk \over \partial k_\alpha}\ \lt(U_{\tilde\phi_\bk} \sigma^y U^\dagger_{\tilde\phi_\bk} \rt)  \no\\
		&=& {1\over 2} \lt( {\partial \tilde\phi_\bk \over \partial k_\alpha}\sigma^z + 
		{\partial \tilde\theta_\bk \over \partial k_\alpha}(\cos\tilde\phi_\bk \sigma^y -\sin\tilde\phi_\bk \sigma^x) \rt) \equiv \mbox{\boldmath{$B_\alpha$}}\cdot \vsigma
	\end{eqnarray}
	where 
	\begin{eqnarray}
		\mbox{\boldmath{$B_x$}} &=& \lt\{ -{1\over 2} {\partial \tilde\theta_\bk \over \partial k_x}\sin\tilde\phi_\bk ,\ {1\over 2} {\partial \tilde\theta_\bk \over \partial k_x}\cos\tilde\phi_\bk, {1\over 2}{\partial \tilde\phi_\bk \over \partial k_x} \rt\} 
		\no\\
		\mbox{\boldmath{$B_y$}} &=& \lt\{ -{1\over 2} {\partial \tilde\theta_\bk \over \partial k_y}\sin\tilde\phi_\bk ,\ {1\over 2} {\partial \tilde\theta_\bk \over \partial k_y}\cos\tilde\phi_\bk, {1\over 2}{\partial \tilde\phi_\bk \over \partial k_y} \rt\}.
	\end{eqnarray}
	Therefore, 
	\begin{eqnarray}
		U_\bk^\dagger{\cal R}_\bk U_\bk &=& U_\bk^\dagger\lt({\partial{\cal C}_\bk \over \partial k_x} {i\partial U_\bk\over \partial k_y}  - {\partial{\cal C}_\bk \over \partial k_y} {i\partial U_\bk \over \partial k_x}\rt)U_\bk \no\\ \no\\
		&=& U_\bk^\dagger\lt[ (\mbox{\boldmath{$A_x$}}\cdot \vsigma)(\mbox{\boldmath{$B_y$}}\cdot \vsigma) - (\mbox{\boldmath{$A_y$}}\cdot \vsigma)(\mbox{\boldmath{$B_x$}}\cdot \vsigma) \rt] U_\bk \no\\
		&=& \lt( \mbox{\boldmath{$A_x$}}\cdot \mbox{\boldmath{$B_y$}} - \mbox{\boldmath{$A_y$}}\cdot \mbox{\boldmath{$B_x$}} \rt) + U_\bk^\dagger i\vsigma\cdot \lt( \mbox{\boldmath{$A_x$}}\times \mbox{\boldmath{$B_y$}} - \mbox{\boldmath{$A_y$}}\times \mbox{\boldmath{$B_x$}} \rt ) U_\bk
	\end{eqnarray}
	The second term is anti-hermitian and should vanish. We have numerically checked that the second term goes to zero for all values of $k$, which yields from Eq.~\ref{UMU},
	\begin{eqnarray}
		\chi_\cmc^\inter &=& {ne^3\over m^2c}\left(1 - {m^2\beta^2\over n} \sum_{\bk i} f(\ek^i)\left( 4k^2 -(U_\bk^\dagger {\cal R}_\bk U_\bk)_{ii} \right)\right) \no\\
		&=&  {ne^3\over m^2c}\lt( 1- 23.0(1)m^2\beta^2n^{2/3}\rt)
		\label{Mag3d}
	\end{eqnarray}
	Therefore, the interband Hall resistivity is given by,
	\begin{eqnarray}
		R^\inter_{\rm H} &\simeq& {1\over nec}  \left( 1-23.0(1) m^2\beta^2 n^{2/3} \right)
	\end{eqnarray}
\clearpage
\end{widetext}

\end{document}